\documentclass[aps,prl,twocolumn,showpacs,superscriptaddress,groupedaddress, secnumarabic,floatfix]{revtex4-1}  

\usepackage{graphicx}  
\usepackage{dcolumn}   
\usepackage{bm}        
\usepackage{amssymb}   
\usepackage{ulem,color}
\usepackage{cancel}
\usepackage{hyperref}
\usepackage[caption=false]{subfig}

\begin{document}

\widetext
\hspace{5.2in} \mbox{MIT-CTP 4317}

%
\affiliation{Massachusetts Institute of Technology, Cambridge, Massachusetts 02139}
\affiliation{Houghton College, Houghton, New York 00000}
\author{Keith Rehermann} \affiliation{Massachusetts Institute of Technology, Cambridge, Massachusetts 02139} 
\author{Christopher M.~Wells} \affiliation{Houghton College, Houghton, New York 14744}
%
%
%
\vskip 0.25cm
  
\DeclareGraphicsRule{.tif}{png}{.png}{`convert #1 `dirname #1`/`basename #1 .tif`.png}

\def\lsim{\mathrel{\rlap{\lower4pt\hbox{\hskip1pt$\sim$}}
    \raise1pt\hbox{$<$}}}
\def\gsim{\mathrel{\rlap{\lower4pt\hbox{\hskip1pt$\sim$}}
    \raise1pt\hbox{$>$}}} 
\newcommand{\vev}[1]{ \left\langle {#1} \right\rangle }
\newcommand{\bra}[1]{ \langle {#1} | }
\newcommand{\ket}[1]{ | {#1} \rangle }

\newcommand{\ev}{ {\rm eV} }
\newcommand{\kev}{{\rm keV}}
\newcommand{\mev}{{\rm MeV}}
\newcommand{\gev}{{\rm GeV}}
\newcommand{\tev}{{\rm TeV}}
\newcommand{\mpl}{$M_{Pl}$}
\newcommand{\mw}{$M_{W}$}
\newcommand{\mz}{$M_{Z}$}

\newcommand{\hc}[1]{#1^{\dagger}}

\newcommand{\Ft}{F_{T}}
\newcommand{\Zparity}{$\mathbb{Z}_2$}
\newcommand{\BLambda}{\boldsymbol{\lambda}}
\newcommand{\Bgamma}{\boldsymbol{\gamma}}
\newcommand{\be}{\begin{eqnarray}}
\newcommand{\ee}{\end{eqnarray}}
\newcommand{\fslash}{\displaystyle{\not}}

\newcommand{\boldgamma}{\textbf{\scriptsize{$\gamma$}}}
\newcommand{\bolde}{\textbf{\scriptsize{e}}}
\newcommand{\rhat}{\hat{\vec{r}}}
\newcommand{\phat}{\hat{\vec{p}}}

\newcommand{\parfrac}[2]{\left(\frac{#1}{#2}\right)}

\newenvironment{frcseries}{\fontfamily{frc}\selectfont}{}
\newcommand{\textfrc}[1]{{\frcseries#1}}
\newcommand{\mathfrc}[1]{\text{\textfrc{#1}}}

\newcommand{\im}{\mathcal{I}m}

\newcommand{\mnu}{m_{\nu}}
\newcommand{\dMnuAtm}{\Delta m^2_{\rm{atm}}}
\newcommand{\dMnuSol}{\Delta m^2_{\rm{\odot}}}

\newcommand{\HubbleAtM}[1]{H|_{T=M_{#1}}}

\newcommand{\nuc}{\nu^c}
\newcommand{\Ynuc}{Y^{\nu^c}}
\newcommand{\Msigma}{M_{\Sigma}}
\newcommand{\SigmaBar}{\overline{\Sigma}}
\newcommand{\order}{\mathcal{O}}
\newcommand{\op}{\mathcal{O}}
\newcommand{\lag}[1]{\mathcal{L_{#1}}}
\newcommand{\Nwidth}{\Gamma_{N_1}}

\newcommand{\KR}[1]{{\bf \color{red} (#1 -k)}}

\newcommand{\tanb}{\tan\beta}
\newcommand{\ssoftmass}[1]{\tilde{m}^2_{#1}}
\newcommand{\bino}{\tilde{B}}
\newcommand{\snu}{\tilde{\nu}}
\newcommand{\stauc}{\tilde{\tau}^c}
\newcommand{\stopc}{\tilde{t}^c}

\newcommand{\mpsi}{m_{\psi}}
\newcommand{\Mpsi}{M_{\psi}}
\newcommand{\mphi}{m_{\phi}}
\newcommand{\Mphi}{M_{\phi}}

\newcommand{\dFtheta}{\int d^4 \theta}
\newcommand{\dTtheta}{\int d^2 \theta}
\newcommand{\Dalpha}[1]{D#1{\alpha}}
\newcommand{\sPot}{\mathcal{W}}
\newcommand{\mgrav}{m_{3/2}}

\newcommand{\eqn}[1]{Eq.\,(\ref{#1})} 
\newcommand{\tab}[1]{Table\,\ref{#1}} 
\newcommand{\fig}[1]{Fig.\,\ref{#1}}
\newcommand{\myref}[1]{Ref.\,\cite{#1}}
\title{Weak Scale Leptogenesis, $R$-symmetry, and a Displaced Higgs}     
\date{\today}

\begin{abstract}
We present a supersymmetric model of leptogenesis in which the right-handed neutrinos have weak scale masses and  $\order(10^{-2}$-$10^{-3})$ yukawa couplings. The model employs an $R$-symmetry at the weak scale that forces neutrino masses to be proportional to the gravitino mass. We predict that the lightest right-handed neutrino decays to a displaced Higgs and neutrino and that the NLSP is long-lived on collider scales. These could be striking signatures at the LHC if, for instance,  the right-handed neutrino is produced in a cascade decay. 
\end{abstract}

\maketitle

\section{Introduction}
The dynamics that generate the observed baryon density remain an unresolved problem.  A number of models have been developed that address the baryon asymmetry, among which leptogenesis \cite{Kuzmin:1985mm,*Fukugita:1986hr,*Luty:1992un} stands out for its economy and simplicity.  In the  simplest version, right-handed neutrinos with a large ($\sim 10^{12} \; \gev$) Majorana mass generate the scale of neutrino masses via the Seesaw mechanism \cite{Yanagida:1980xy,*1980PhRvL..44..912M} and have $CP$ violating decays which result in a lepton asymmetry.  
Despite their simplicity, such dynamics have little hope of ever being observed.\\
\indent In this letter we argue that leptogenesis with weak scale right-handed neutrinos \cite{Flanz:1996fb,*Covi:1996wh,*Pilaftsis:1997jf,*Pilaftsis:2003gt, Blanchet:2010kw,Haba:2011ra} can be achieved if Supersymmetry (SUSY) breaking is mediated in an $R$-symmetric fashion. Tying together $R$-symmetric SUSY breaking and weak scale leptogenesis leads to a number of predictions. 
First, the neutrino masses are controlled by the gravitino mass. Second, successful leptogenesis predicts that the decays of the lighest right-handed neutrino produce a displaced Higgs and neutrino. This is a rather novel signature.  Third, the NLSP is long-lived on collider scales. Together these lead to an intriguing connection between SUSY breaking, Electroweak Symmetry Breaking (EWSB) and neutrino masses, while providing a unique LHC signature in events with a right-handed neutrino.\\ 
\indent Turning an eye toward model building, we have two more interesting observations. First, the $R$-symmetry allows the hypothesized dimensionless couplings in the low energy theory to be $\order(10^{-2}$-$10^{-3})$,  avoiding the more serious tunings that often appear in other attempts to model weak scale leptogenesis. Second, since the dynamics of leptogenesis are at the weak scale, the gravitino problem  \cite{Krauss:1983ik,*Cheung:2011mg} that plagues SUSY models with high scale right-handed neutrinos is easily avoided. 
\vspace{0.41in}
\section{Model}
Our model is based on three hypotheses.  First, we take the superpotential to be that of the MRSSM \cite{Kribs:2007ac} extended to include right-handed neutrinos
\be
\label{eq:sup}
\sPot \supset \mu_u H_u R_u + \mu_d H_d R_d + y_{N_{ij}} R_d L_i N_j +\frac{1}{2}M_{ij}N_i N_j. \;\;\;\;\;
\ee
The fields $R_{u,d}$ are weak doublets with hypercharge opposite the corresponding MSSM Higgs and $N_j$ are three generations of right-handed neutrinos.  The right-handed neutrino mass term is consistent with a $\mathbb{Z}_4^R$ $R$-symmetry \cite{Lee:2010gv,*Lee:2011dya}; charge assignments are shown in Table \ref{tab:z2}. 
Our second assumption is that $D$ and $F$ term SUSY breaking occur at a common scale. 
Finally, we assume that the dominant $R$-violation from SUSY breaking  is the irreducible contribution from Supergravity (SUGRA) 


\be
\label{eq:Lrviolation}
\delta L_{\not{R}} \supset \dTtheta
\Phi (\mu_d H_d R_d + \mu_u H_u R_u)  \nonumber \\ = \mgrav(\mu_d H_d R_d + \mu_u H_u R_u + \textrm{c.c.})
\ee
where $\Phi=1+\mgrav \theta^2$ is the conformal compensator  \cite{Siegel:1978mj,*Gates:1983nr,*Kaplunovsky:1994fg}.

After EWSB, our assumptions lead to vevs for the scalars $r^0_d, \; r^0_u$ 
\be
\label{eq:rvevs}
\vev{r^0_d}=\frac{\mgrav \mu_d v_d}{\mu^2_d+m^2_{soft}} \; \; , \; \; \vev{r^0_u}=\frac{\mgrav \mu_u v_u}{\mu^2_u+m^2_{soft}}.
\ee
We see that the irreducible $R$-violation induced by SUGRA puts the $r$'s at vevs proportional to $\mgrav$. Consequently, the left-handed neutrinos have Majorana masses controlled by $\mgrav^2/M_N$.  We note, in passing, that if one enforces an exact 
$U(1)_R$ the Majorana masses in \eqn{eq:sup} are forbidden. In this case the neutrinos are Dirac and $\mnu \propto \mgrav$.  This is an interesting possibility which would imply low scale SUSY breaking
\footnote{KR thanks Matthew McCullough for an insightful discussion on this point}.

Our hypotheses necessitate Dirac masses for the gauginos \cite{Hall:1990hq,*Randall:1992cq,*Fox:2002bu}. We remain agnostic concerning the UV origin of these masses, parameterizing them as  
\be
\label{eq:diracmasses}
\int d^2\theta\frac{\theta_{\alpha} D'}{\Lambda} W_i^{\alpha} \Phi_i,
\ee
where $\theta_{\alpha} D'$ a is D-type spurion with $R$-charge $+1$ and the index $i$ runs over the SM gauge groups.  The $\Phi_i$ are adjoint chiral superfields whose fermionic components $\psi_i$ marry the gauginos through Dirac mass terms $\propto (D'/\Lambda)\lambda_i\psi_i$.  Weak scale soft masses can be generated if, for example,  $D' \!\sim\! F_X\! \sim\! 10^{15}\,\gev^2$ and $\Lambda \! \sim\! 10^{12}\,\gev$. Studies of $R$-symmetric mediation suited to our model can be found in Ref's.\,\cite{Amigo:2008rc,Blechman:2009if,Benakli:2010gi}.

We note that the $\mu$ terms in \eqn{eq:sup} need not be explicit. Rather, they can be generated by the Giudice-Masiero mechanism \cite{1988PhLB..206..480G} if we assume the existence of an additional $\mathbb{Z}_2$ symmetry, as in \tab{tab:z2} \cite{Davies:2011mp}. 
\begin{table}
\begin{center}
    \begin{tabular}{  | l  | c  l  r  |  }
            \hline
   Field & $ \mathbb{Z}_4^R$ & \Zparity &\\ \hline\hline
    $H_{u,d}$ & $0$  & 0&\\ 
     $N$ & $-1$ &1&\\ 
      $R_{u,d}, X$ & $2$  & 1&\\ 
    $Q,u^c,d^c, L, e^c$ & $1$ & 0& \\ \hline

  \end{tabular}
\end{center}
 \caption{\label{tab:z2} $R$-charges and parity assignments that can generate the superpotential in \eqn{eq:sup} once $F_X \neq 0$.}

\end{table}
These charge assignments allow the following operators up to dim 6 
\be
\label{eq:higher_dim}
\lag{}\!\supset \!\! \dFtheta \frac{\hc{X}}{\Lambda}\sum_{i=u,d}c_i R_i H_i  + c_3\frac{X\hc{X}}{\Lambda^2}H_uH_d +\textrm{h.c.}
\ee
For $c_1,c_2,c_3$ of $\order(1)$ and $\frac{F_X}{\Lambda}\! \sim\! m_{\textrm{\scriptsize{soft}}}$ these operators generate the $\mu_u,\mu_d,b_h$ parameters that trigger EWSB.  

Conversely, given the field content in \tab{tab:z2} $M_N$ cannot be generated in an analogous manner. In this sense $M_N$ is a free parameter, though as we shall see, successful leptogenesis requires $M_N \lsim \order(\tev)$. It would be interesting to dynamically explain this scale \footnote{One possibility would be if there existed $X'$ with $R$=-2 and $F_{X'} \sim F_{X}$}.


\section{Leptogenesis, Neutrino Masses and $R$-symmetry}
We briefly review the salient features of leptogenesis; see \myref{Davidson:2008bu} for a thorough review. Ultimately the aim of leptogenesis is to explain the {\it baryon} asymmetry
\be
\label{eq:YdB}
Y_{\Delta b} \equiv \frac{n_b-n_{\bar{b}}}{s} \approx 9 \times 10^{-11}
\ee
where $n_{b}$, $n_{\bar{b}}$ are the number densities of baryons and anti-baryons respectively, $s$ is the entropy density and $Y_{\Delta b}$ is referred to as the baryon asymmetry yield. In the early universe an asymmetry in lepton number can be converted to an asymmetry in baryon number by sphaleron processes. Schematically we have
\be
Y_{\Delta b} \simeq C_{\rm sph} Y_{\Delta l}  
\ee
where $Y_{\Delta l}$ is defined in analogy to $Y_{\Delta b}$ and typically $1/2 \gsim C_{\rm sph} \gsim 1/4$.

A non-zero $Y_{\Delta l} $ can be generated if the Sakharov conditions (lepton number violation,  $CP$ violation and out of equilibrium dynamics) \cite{Sakharov:1967dj} are satisfied. The lepton asymmetry yield can be parameterized by the degree of $CP$ violation $\epsilon$ and the efficiency of out of equilibrium dynamics $\eta$:
\be
\label{eq:YdL}
Y_{\Delta l} \simeq \frac{k}{g_*} \epsilon \eta
\ee
where $g_*$ is the number degrees of freedom in equilibrium, typically $\order(100)$, and $k$ is an $\mathcal{O}(1)$ factor.  Note that by definition $\epsilon, \; \eta < 1$. 

We now turn our attention to the interplay between neutrino masses and $Y_{\Delta l}$ in the model presented above. Left-handed neutrino masses are obtained from \eqn{eq:sup} by integrating out the right-handed Majorana neutrinos:
\be
\label{eq:nu_masses}
(m_{\nu})_{ij}  = \sum_k \frac{1}{M_{N_k}} (y_N y_N^T)_{ij}v^2_{r_d} 
\ee
in the basis that diagonalizes $M_{ij}$.  As a direct consequence of the $R$-symmetry,  Eq's.\,(\ref{eq:rvevs}) and (\ref{eq:nu_masses}) imply that $\mnu \propto \mgrav^2/M_N$.  Said differently, the $R$-symmetry protects $\vev{r^0_d}$ thereby allowing weak scale $M_N$ without tuning the scalar potential.

\begin{figure}[H,t]
\begin{centering}
\subfloat[]{\raisebox{1mm}{{\label{fig:wavefunction}}\includegraphics[width=0.25\textwidth]{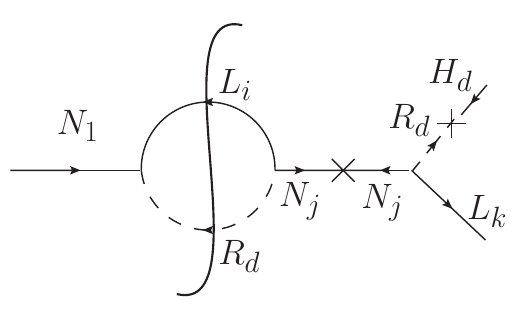}}} \hfill
\subfloat[]{{\label{fig:triangle}}\includegraphics[width=0.23\textwidth]{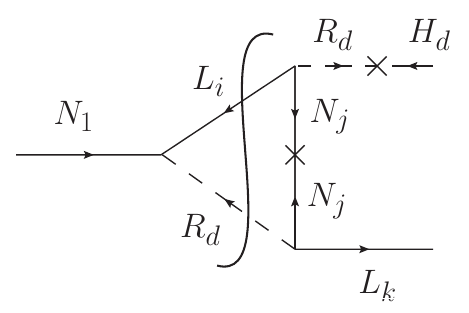}}   \\
\subfloat[]{{\label{fig:wash1}}\includegraphics[width=0.2\textwidth]{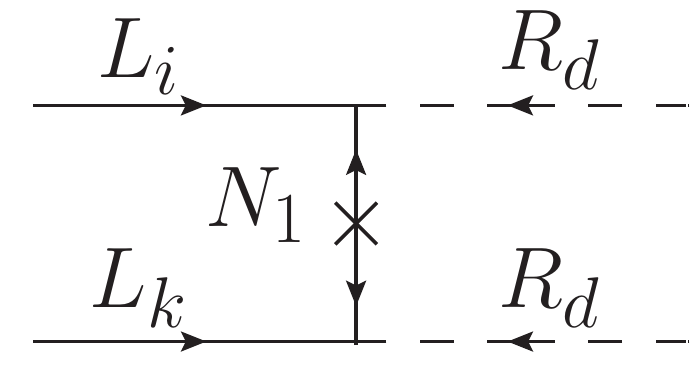}} \hfill
\subfloat[]{{\label{fig:wash2}}\includegraphics[width=0.2\textwidth]{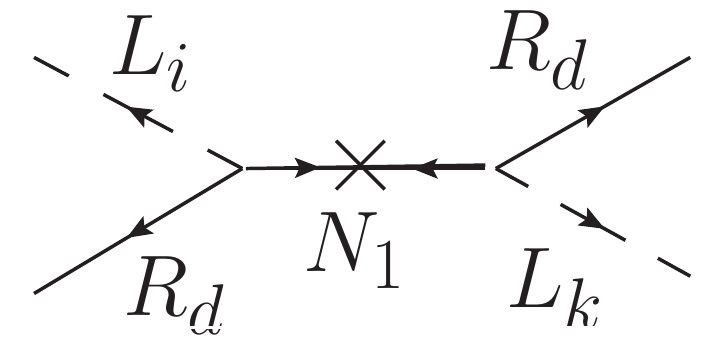}}
\par\end{centering}
\caption{\label{fig:washout_diagrams}
The top row illustrates the diagrams contributing to $\epsilon$. 
The bottom shows some of the diagrams that contribute to washout and suppress $\eta$.}
\end{figure}

The degree of $CP$ violation is governed by the parameter
\be
\label{eq:epsilon}
\epsilon = \frac{\Gamma(N\rightarrow \nu h) - \Gamma(N\rightarrow \hc{(\nu h)})}{\Gamma(N\rightarrow \nu h) + \Gamma(N\rightarrow \hc{(\nu h)})}.
\ee
where $h$ is the physical Higgs \footnote{The next section explains why it is the physical Higgs.}. Interference between tree and one loop diagrams, examples of which are shown in Figures \ref{fig:wavefunction} and \ref{fig:triangle}, generate $\epsilon$:
\be
\label{eq:epsilonApp}
\!\!\epsilon \approx -\frac{3}{16\pi} \frac{1}{(\hc{y}_{N} y_{N})_{11}} \sum_j \im\left[\left((\hc{y}_Ny_{N})_{1j}\right)^2\right] \frac{M_{N_1}}{M_{N_j}}. \; \;
\ee
This is an adequate approximation for our purposes if $M_{N_{2,3}} > 3 M_{N_1}$ and we take this to be the case in what follows.
The scaling of $\epsilon$ is more obvious if we
make the simplifying assumption $y_{N_{ij}} \approx y_{N_i}$ ({\it i.e.}\,entries along a particular row are comparable)
\be
\label{eq:simpleEps}
|\epsilon| &\approx& \frac{3}{16\pi} \sum_j |\hc{y}_{N_j} y_{N_j}| \sin\theta_j \frac{M_{N_1}}{M_{N_j}} \simeq \frac{3M_{N_1}}{16\pi v^2_{r_d}}   m^{\rm max}_{\nu} \sin\delta \nonumber \\
&\simeq& 10^{-3} \parfrac{M_{N_1}}{300 \; \gev}\!\!\parfrac{10^{-3} \; \gev}{v_{r_d}}^2\!\!\parfrac{m^{\rm max}_{\nu}}{0.1 \; \ev}\!\sin\delta\: \: \; \;\;\;\;
\ee
%
where $m^{\rm max}_{\nu}$ is the largest left-handed neutrino mass and $\sin \delta$ is the associated $CP$ violating phase. 
The scaling from \eqn{eq:rvevs},  $\vev{r^0_d}\!\sim\!\mgrav$, allows a large 
$CP$ asymmetry to be generated for a weak scale right-handed neutrino, yukawa couplings of $\order(10^{-2}$-$10^{-3})$ and $\mgrav$ of $ \order(1 \; \mev)$. Note that the scale of $R$-symmetry  violation simultaneously determines the neutrino masses and the magnitude of $CP$ violation.

We now estimate the efficiency $\eta$ of the out of equilibrium dynamics.  Our goal in this letter is not to map the entire parameter space of viable leptogenesis, but to show that such a parameter space exists and that it is sizable. To this end we work in the ``weak-washout" regime, $\eta$ of $\order(1)$. This regime applies if $N_1$ decays out of equilibrium and the rates for all other lepton number violating processes (such as inverse decay and $2\rightarrow 2$ scattering) are small compared to the Hubble rate. In this approximation  the asymmetry is set by $\epsilon$.  While successful leptogenesis may occur outside of the weak-washout regime - due to a delicate interplay between the dynamics contributing to $\epsilon$ and $\eta$ -  this simplifying assumption is sufficient for our goal and in what follows we determine the parameter space that satisfies $\eta$ of $\order(1)$.  

The out of equilibrium condition requires that the decay rate of $N_1$ is less than the Hubble rate at temperatures near $M_{N_1}$,
\be
\label{eq:gammaBound}
\Gamma < g_* \frac{M^2_{N_1}}{M_{pl}} \simeq 10^{-12} {\rm sec}^{-1} \parfrac{M_{N_1}}{300 \;  \gev}^2.
\ee
It is generically difficult to achieve this for a weak scale mass because it implies that the width is many orders of magnitude smaller than the mass, $\Gamma_{N_1}/M_{N_1}\lsim 10^{-12}(M_{N_1}/300 \; \gev)$. However, our model naturally suppresses processes that violate the $R$-symmetry. In fact, the width $N_1$ is protected by the $R$-symmetry if it is lighter than the scalar $r^0_d$.  In this case, it decays through the $R$-violating channel $N_1\rightarrow \nu h$, where $h$ is the physical Higgs. As with neutrino masses and the magnitude of $CP$ violation, the scale of $R$-violation sets this decay rate: 
\be
\!\!\Gamma
&=& \frac{(\hc{y_{N}} y_N)_{11}}{8 \pi}  \frac{b_d}{m^2_{r_d}} M_{N_1} = \frac{(\hc{y_N} y_N)_{11}}{8 \pi} \left(\frac{v_{r_d}}{v_{h_d}}\right)^2 M_{N_1} \; \; \;
\ee
where the second expression follows directly from the tadpole equation for $r^0_d$, \eqn{eq:rvevs}.
The time scale in \eqn{eq:gammaBound} suggests that $N_1$ decays are displaced on collider scales. We revisit this point in the next section. 

Lepton number violating processes, such as those illustrated in \fig{fig:wash1} and \fig{fig:wash2}, can also washout the asymmetry, suppressing $\eta$. However,  if all scalars are heavy except for a light Higgs, then the washout processes decouple because their rates are Boltzmann suppressed. Assuming that the left handed sleptons and $r^0_d$ are heavy with common mass $\tilde{M}$, we can estimate the ratio $\tilde{M}/M_N$ necessary to suppress washout.  In analogy to \eqn{eq:gammaBound}, the washout rate must be less than the Hubble rate.
The rate for non-relativistic scalar scattering with a relativistic fermion is \cite{Kolb:1990vq}
\be
\label{eq:rate}
\!\!\!\!\Gamma_{\textrm{\scriptsize{Washout}}}\equiv n\vev{\sigma v} \approx (\tilde{M}T)^{3/2} {\rm e}^{-\tilde{M}/M_N} \frac{\alpha^2_y}{ (s-M^2_N)} 
\ee
where $\alpha_y=y^2/(4\pi)$.
%
%


\begin{figure}
\includegraphics[width=0.5\textwidth]{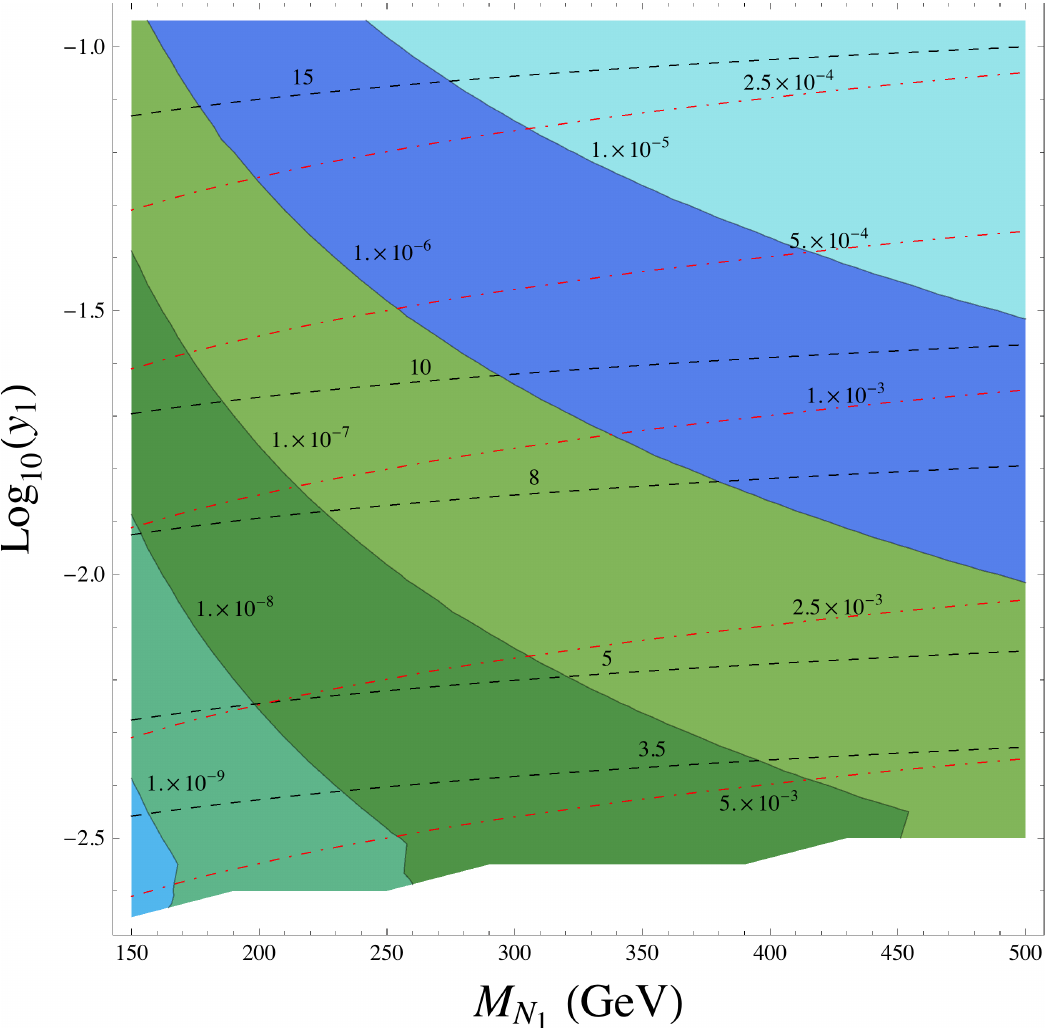}
\caption{This plot illustrates that a significant parameter space exists for successful leptogenesis, even with the simplifying assumptions in the text. Shaded regions are $Y_{\Delta l}$ at $T=100 \; \gev$, assuming maximal $CP$ violation. Red dashed-dotted line are $\vev{r^0_d} \; (\gev)$. Black dashed lines are values of $\tilde{M}/M_N$ that result in suppressed washout, per Eq.\,(\ref{eq:rate}).}
\label{fig:money_plot}
\end{figure}

To determine the lepton asymmetry yield 
we solve the following Boltzmann equation
\be
\label{eq:boltz}
\dot{Y}_{\Delta l} &=& \epsilon \Gamma_{\textrm{\scriptsize{Decay}}}(Y_{N_1}-Y^{\textrm{\scriptsize{eq}}}_{N_1}) - \Gamma_{\textrm{\scriptsize{Washout}}}Y_{\Delta l}
\ee
where we have defined
\be
\dot{Y} &\equiv& \frac{s H|_{T=M_{N_1}}}{z} \frac{dY}{dz}, \qquad z \equiv \frac{M_{N_1}}{T}.
\ee
Figure \ref{fig:money_plot} plots solutions to \eqn{eq:boltz} where we assume maximal $CP$ violation in \eqn{eq:epsilonApp}, fix $v_{h_d}= v/10$ ($\tan\beta =10$) and enforce \eqn{eq:gammaBound}.
We also plot contours of constant $\vev{r^0_d}$ and $\tilde{M}/M_{N_1}$, while omitting the region $\tilde{M}/M_{N_1}<3$ since our approximation in \eqn{eq:rate} is not valid in this regime.  The plot illustrates that, even with our simplifying assumptions, a sufficient asymmetry is generated in a large region of parameter space. The entire range in \fig{fig:money_plot} allows a phenomenologically viable value $Y_{\Delta l} \sim 10^{-10}$ by moving away from the assumption of maximal $CP$ violation and $\eta$ of $\order(1)$.

We note that there exists some viable parameter space in the regime $\tilde{M} \lsim M_{N_1}$. In this case only the smallness of $y_N$ can prevent washout, which in turn implies smaller $\epsilon$.  We estimate the constraint on $y_N$ by using the relativistic form of the rate in \eqn{eq:rate} and by requiring it to be less than the Hubble rate at $T=M_{N_1}$
\be
\label{eq:relate}
y_N\lsim 10^{-3} \left(\frac{g_*}{100}\frac{M_{N_1}}{300 \; \gev}\right)^{1/4} \Rightarrow \epsilon \lsim 10^{-7}
\ee
where the $\epsilon$ bound follows from \eqn{eq:epsilonApp}. 

Since there exist viable solutions in both regimes, we conclude that the interpolating regime of moderate yukawa couplings and comparable masses offers the possibility that the yukawa couplings and the mass ratio need not be tuned against each other to the extent in our simple estimates. We leave a more detailed exploration of this regime for later work.


\section{A Displaced Higgs at the LHC?}

A spectrum that is consistent with our assumptions is heavy gauginos and scalar superpartners $ \tilde{M}\!\sim\!\tev$, moderate right-handed neutrinos and higgsinos $ M_{N_1} \sim \mu_{u,d}$ and a light Higgs. 
In this regime the lightest right-handed neutrino decays to the Higgs and a neutrino. The out of equilibrium condition in \eqn{eq:gammaBound} sets the lifetime of the lightest right-handed neutrino.  Intriguingly, this condition {\it necessitates} that $N_{1} \rightarrow \nu h$ is a displaced decay. This follows from demanding that $N_{1}$ has a weak scale mass and that  it decays
before the sphaleron processes turn off at $T \sim 100\,\gev$. The lifetime of $N_1$ is then in the range
\be
\left(H|_{T=100 \; \gev}\right)^{-1} \gsim \tau_{N_1} \gsim \left(\HubbleAtM{N_1}\right)^{-1} \Rightarrow \nonumber\\ \sim 10^{-11} \; {\rm sec} \gsim  \tau_{N_1} \gsim 10^{-12}\left(\frac{300 \; \gev}{M_{N_1}}\right)^2 {\rm sec} 
\ee
This lifetime is comparable to those of B and D mesons if $M_{N_1} \lsim \tev$.  ATLAS, CMS, and LHCb all have the ability to resolve decays on this time scale. Therefore, our model has the striking possibility of being observed via displaced Higgs events.

We note that although the direct production cross section of $N_1$ is vanishingly small at the LHC, in the spectrum noted above the sneutrino width can be dominated by $\snu_i \rightarrow N_1 \tilde{r}^0_d$. This occurs if the yukawa couplings $y_N$ are larger than the yukawa couplings in the operator $y_L H_d L e^c$. Generating a sufficient asymmetry requires $y_N \gsim 10^{-4}$, and at minimum we expect $\snu_e$ to dominantly decay through $\snu_{e}\rightarrow N_1 \tilde{r}^0_d$. Therefore, cascade decays involving $\snu_e$ terminate with a displaced Higgs. The same may be true for the other flavors depending on the exact values of the  yukawa couplings.
%
%

Finally, we draw attention to the fact that successful leptogenesis puts the gravitino mass at $\order(1\; \mev)$.  Therefore the NLSP, $\chi$,  is long-lived on collider scales:
 \be
 \tau_{\chi} &\approx& 16 \pi \frac{F^2}{m^5_{\chi}} \approx 10^{-2} \; {\rm s} \;  \parfrac{\mgrav}{ 1 \; \mev}^2 \left(\frac{100 \gev}{\tilde{m}_{\chi}}\right)^5. \;\;\;
\ee
where we have neglected phase space suppression. The NLSP must then appear either as missing energy, a CHAMP \cite{Dine:1995ag,*Dine:1994vc,*Feng:1999fu}, or an $R$-hadron \cite{Raby:1997pb,*Baer:1998pg,*Giudice:2004tc,*ArkaniHamed:2004fb}.  This gives us an additional distinctive collider signature accompanying the prediction of a displaced Higgs. We intend to explore these possibilities in future work.

\section{A Note on Dirac Gauginos}

For the sake of completeness, we recall a few well-known complications to the minimal Dirac gaugino story and simple extensions that address them.  The hypercharge adjoint $\Phi_Y$, which gives mass to the Bino, is a complete SM and $\mathbb{Z}_4^R$ singlet.  As such, our effective theory should contain terms like 
\be
\label{eq:phiY}
\mathcal{L} \supset \int d^4\theta \frac{\hc{X}X}{\Lambda}\Phi_Y + \int d^2\theta\theta^\alpha\theta_\alpha \frac{D'^2}{\Lambda} \Phi_Y +\,\textrm{h.c.}.
\ee 
If $\Phi_Y$ receives a soft mass $\tilde{m}_{\textrm{\scriptsize{soft}}}^2$, then the tadpole above pushes $\Phi_Y$ to a vev on the order of 
\be
\label{eq:vevphiY}
|\vev{\Phi_Y}| \sim \frac{F_X^2+D'^2}{\Lambda\,\tilde{m}_{\textrm{\scriptsize{soft}}}^2} \sim \frac{F_X^2}{\Lambda \left(\frac{F_X}{\Lambda}\right)^2} = \Lambda.
\ee
This cutoff scale vev introduces a D-term for hypercharge of order $D'$ via Eq.\,(\ref{eq:diracmasses}).  Also, because $\Phi_Y$ is an $R$-symmetry singlet, its vev introduces cutoff size corrections to $\mu_{u,d}$.  Both of these effects can spoil EWSB.  Furthermore, singlet couplings to the supergravity multiplet generically reintroduce quadratic divergences and destabilize the hierarchy \cite{Bagger:1995ay}.  

There are a number of UV solutions to these problems compatible with our low energy phenomenology:
\myref{Amigo:2008rc} utilizes an unbroken discrete symmetry, \myref{Benakli:2010gi} exploits a clever set of couplings between the messenger fields and the adjoints and \myref{Abel:2011dc} forbids these terms by embedding the theory into $SU(5)$. 
\section{Discussion}

We have presented a model of leptogenesis which is connected to the physics of the weak scale through SUSY breaking and $R$-symmetry.  The dynamics of neutrino masses and leptogenesis are both controlled by the scale of $R$-violation in the low energy theory.  The model ties the neutrino mass scale to the gravitino mass, which in conjunction with demanding a large enough lepton asymmetry, predicts $\mgrav$ of $\order(\mev)$.  Furthermore, there are striking LHC signatures: cascade decays through the lightest right-handed neutrino terminate with a displaced Higgs and the NLSP is long-lived on collider scales.  Finally, the gravitino problem is avoided because all of the interesting visible sector dynamics are at the weak scale which allows for reheating temperatures of $\order(\tev)$.
\\
\indent Of course, nothing is free, and our dynamics rely on $R$-symmetric mediation of SUSY breaking.  This introduces hurdles to, for example, unification.  However, we find the interplay of supersymmetry breaking, neutrino masses and the potential observability of leptogenesis dynamics at the LHC a tantalizing possibility. There are a number of directions for further study including a refined numerical study of the dependence of the asymmetry on the spectrum of soft masses, a dedicated study of the collider phenomenology, and more detailed considerations of UV model building. We intend to pursue these topics in future publications.

\section{Acknowledgments}
\begin{acknowledgments}
We thank Matthew McCollough and Jesse Thaler for insightful conversions and Matthew McCollough, David E. Kaplan, and Gordan Krnjaic for comments on the manuscript. This work is supported by the U.S. Department of Energy under cooperative research agreement DE-FG02-05ER-41360.  CMW was partially supported by the Houghton College Summer Research Institute.
\end{acknowledgments}


\begin{thebibliography}{50}%
\bibliographystyle{apsrev4}
\makeatletter
\providecommand \@ifxundefined [1]{%
 \@ifx{#1\undefined}
}%
\providecommand \@ifnum [1]{%
 \ifnum #1\expandafter \@firstoftwo
 \else \expandafter \@secondoftwo
 \fi
}%
\providecommand \@ifx [1]{%
 \ifx #1\expandafter \@firstoftwo
 \else \expandafter \@secondoftwo
 \fi
}%
\providecommand \natexlab [1]{#1}%
\providecommand \enquote  [1]{``#1''}%
\providecommand \bibnamefont  [1]{#1}%
\providecommand \bibfnamefont [1]{#1}%
\providecommand \citenamefont [1]{#1}%
\providecommand \href@noop [0]{\@secondoftwo}%
\providecommand \href [0]{\begingroup \@sanitize@url \@href}%
\providecommand \@href[1]{\@@startlink{#1}\@@href}%
\providecommand \@@href[1]{\endgroup#1\@@endlink}%
\providecommand \@sanitize@url [0]{\catcode `\\12\catcode `\$12\catcode
  `\&12\catcode `\#12\catcode `\^12\catcode `\_12\catcode `\%12\relax}%
\providecommand \@@startlink[1]{}%
\providecommand \@@endlink[0]{}%
\providecommand \url  [0]{\begingroup\@sanitize@url \@url }%
\providecommand \@url [1]{\endgroup\@href {#1}{\urlprefix }}%
\providecommand \urlprefix  [0]{URL }%
\providecommand \Eprint [0]{\href }%
\@ifxundefined \urlstyle {%
  \providecommand \doi  [0]{\begingroup \@sanitize@url \@doi}%
  \providecommand \@doi [1]{\endgroup \@@startlink {\doibase
  #1}doi:\discretionary {}{}{}#1\@@endlink }%
}{%
  \providecommand \doi  [0]{doi:\discretionary{}{}{}\begingroup
  \urlstyle{rm}\Url }%
}%
\providecommand \doibase [0]{http://dx.doi.org/}%
\providecommand \Doi [0]{\begingroup \@sanitize@url \@Doi }%
\providecommand \@Doi  [1]{\endgroup\@@startlink{\doibase#1}\@@Doi}%
\providecommand \@@Doi [1]{#1\@@endlink}%
\providecommand \selectlanguage [0]{\@gobble}%
\providecommand \bibinfo  [0]{\@secondoftwo}%
\providecommand \bibfield  [0]{\@secondoftwo}%
\providecommand \translation [1]{[#1]}%
\providecommand \BibitemOpen [0]{}%
\providecommand \bibitemStop [0]{}%
\providecommand \bibitemNoStop [0]{.\EOS\space}%
\providecommand \EOS [0]{\spacefactor3000\relax}%
\providecommand \BibitemShut  [1]{\csname bibitem#1\endcsname}%
\bibitem [{\citenamefont {Kuzmin}\ \emph {et~al.}(1985)\citenamefont {Kuzmin},
  \citenamefont {Rubakov},\ and\ \citenamefont {Shaposhnikov}}]{Kuzmin:1985mm}%
  \BibitemOpen
  \bibfield  {author} {\bibinfo {author} {\bibfnamefont {V.}~\bibnamefont
  {Kuzmin}}, \bibinfo {author} {\bibfnamefont {V.}~\bibnamefont {Rubakov}}, \
  and\ \bibinfo {author} {\bibfnamefont {M.}~\bibnamefont {Shaposhnikov}},\
  }\Doi {10.1016/0370-2693(85)91028-7} {\bibfield  {journal} {\bibinfo
  {journal} {Phys.Lett.},\ }\textbf {\bibinfo {volume} {B155}},\ \bibinfo
  {pages} {36} (\bibinfo {year} {1985})}\BibitemShut {NoStop}%
\bibitem [{\citenamefont {Fukugita}\ and\ \citenamefont
  {Yanagida}(1986)}]{Fukugita:1986hr}%
  \BibitemOpen
  \bibfield  {author} {\bibinfo {author} {\bibfnamefont {M.}~\bibnamefont
  {Fukugita}}\ and\ \bibinfo {author} {\bibfnamefont {T.}~\bibnamefont
  {Yanagida}},\ }\Doi {10.1016/0370-2693(86)91126-3} {\bibfield  {journal}
  {\bibinfo  {journal} {Phys.Lett.},\ }\textbf {\bibinfo {volume} {B174}},\
  \bibinfo {pages} {45} (\bibinfo {year} {1986})}\BibitemShut {NoStop}%
\bibitem [{\citenamefont {Luty}(1992)}]{Luty:1992un}%
  \BibitemOpen
  \bibfield  {author} {\bibinfo {author} {\bibfnamefont {M.}~\bibnamefont
  {Luty}},\ }\Doi {10.1103/PhysRevD.45.455} {\bibfield  {journal} {\bibinfo
  {journal} {Phys.Rev.},\ }\textbf {\bibinfo {volume} {D45}},\ \bibinfo {pages}
  {455} (\bibinfo {year} {1992})}\BibitemShut {NoStop}%
\bibitem [{\citenamefont {Yanagida}(1980)}]{Yanagida:1980xy}%
  \BibitemOpen
  \bibfield  {author} {\bibinfo {author} {\bibfnamefont {T.}~\bibnamefont
  {Yanagida}},\ }\href@noop {} {\bibfield  {journal} {\bibinfo  {journal}
  {Prog.Theor.Phys.},\ }\textbf {\bibinfo {volume} {64}},\ \bibinfo {pages}
  {1103} (\bibinfo {year} {1980})}\BibitemShut {NoStop}%
\bibitem [{\citenamefont {{Mohapatra}}\ and\ \citenamefont
  {{Senjanovic}}(1980)}]{1980PhRvL..44..912M}%
  \BibitemOpen
  \bibfield  {author} {\bibinfo {author} {\bibfnamefont {R.~N.}\ \bibnamefont
  {{Mohapatra}}}\ and\ \bibinfo {author} {\bibfnamefont {G.}~\bibnamefont
  {{Senjanovic}}},\ }\Doi {10.1103/PhysRevLett.44.912} {\bibfield  {journal}
  {\bibinfo  {journal} {Physical Review Letters},\ }\textbf {\bibinfo {volume}
  {44}},\ \bibinfo {pages} {912} (\bibinfo {year} {1980})}\BibitemShut
  {NoStop}%
\bibitem [{\citenamefont {Flanz}\ \emph {et~al.}(1996)\citenamefont {Flanz},
  \citenamefont {Paschos}, \citenamefont {Sarkar},\ and\ \citenamefont
  {Weiss}}]{Flanz:1996fb}%
  \BibitemOpen
  \bibfield  {author} {\bibinfo {author} {\bibfnamefont {M.}~\bibnamefont
  {Flanz}}, \bibinfo {author} {\bibfnamefont {E.~A.}\ \bibnamefont {Paschos}},
  \bibinfo {author} {\bibfnamefont {U.}~\bibnamefont {Sarkar}}, \ and\ \bibinfo
  {author} {\bibfnamefont {J.}~\bibnamefont {Weiss}},\ }\Doi
  {10.1016/S0370-2693(96)01337-8} {\bibfield  {journal} {\bibinfo  {journal}
  {Phys.Lett.},\ }\textbf {\bibinfo {volume} {B389}},\ \bibinfo {pages} {693}
  (\bibinfo {year} {1996})},\ \Eprint {http://arxiv.org/abs/hep-ph/9607310}
  {arXiv:hep-ph/9607310 [hep-ph]} \BibitemShut {NoStop}%
\bibitem [{\citenamefont {Covi}\ \emph {et~al.}(1996)\citenamefont {Covi},
  \citenamefont {Roulet},\ and\ \citenamefont {Vissani}}]{Covi:1996wh}%
  \BibitemOpen
  \bibfield  {author} {\bibinfo {author} {\bibfnamefont {L.}~\bibnamefont
  {Covi}}, \bibinfo {author} {\bibfnamefont {E.}~\bibnamefont {Roulet}}, \ and\
  \bibinfo {author} {\bibfnamefont {F.}~\bibnamefont {Vissani}},\ }\Doi
  {10.1016/0370-2693(96)00817-9} {\bibfield  {journal} {\bibinfo  {journal}
  {Phys.Lett.},\ }\textbf {\bibinfo {volume} {B384}},\ \bibinfo {pages} {169}
  (\bibinfo {year} {1996})},\ \Eprint {http://arxiv.org/abs/hep-ph/9605319}
  {arXiv:hep-ph/9605319 [hep-ph]} \BibitemShut {NoStop}%
\bibitem [{\citenamefont {Pilaftsis}(1997)}]{Pilaftsis:1997jf}%
  \BibitemOpen
  \bibfield  {author} {\bibinfo {author} {\bibfnamefont {A.}~\bibnamefont
  {Pilaftsis}},\ }\Doi {10.1103/PhysRevD.56.5431} {\bibfield  {journal}
  {\bibinfo  {journal} {Phys.Rev.},\ }\textbf {\bibinfo {volume} {D56}},\
  \bibinfo {pages} {5431} (\bibinfo {year} {1997})},\ \Eprint
  {http://arxiv.org/abs/hep-ph/9707235} {arXiv:hep-ph/9707235 [hep-ph]}
  \BibitemShut {NoStop}%
\bibitem [{\citenamefont {Pilaftsis}\ and\ \citenamefont
  {Underwood}(2004)}]{Pilaftsis:2003gt}%
  \BibitemOpen
  \bibfield  {author} {\bibinfo {author} {\bibfnamefont {A.}~\bibnamefont
  {Pilaftsis}}\ and\ \bibinfo {author} {\bibfnamefont {T.~E.}\ \bibnamefont
  {Underwood}},\ }\Doi {10.1016/j.nuclphysb.2004.05.029} {\bibfield  {journal}
  {\bibinfo  {journal} {Nucl.Phys.},\ }\textbf {\bibinfo {volume} {B692}},\
  \bibinfo {pages} {303} (\bibinfo {year} {2004})},\ \Eprint
  {http://arxiv.org/abs/hep-ph/0309342} {arXiv:hep-ph/0309342 [hep-ph]}
  \BibitemShut {NoStop}%
\bibitem [{\citenamefont {Blanchet}\ \emph {et~al.}(2010)\citenamefont
  {Blanchet}, \citenamefont {Dev},\ and\ \citenamefont
  {Mohapatra}}]{Blanchet:2010kw}%
  \BibitemOpen
  \bibfield  {author} {\bibinfo {author} {\bibfnamefont {S.}~\bibnamefont
  {Blanchet}}, \bibinfo {author} {\bibfnamefont {P.}~\bibnamefont {Dev}}, \
  and\ \bibinfo {author} {\bibfnamefont {R.}~\bibnamefont {Mohapatra}},\ }\Doi
  {10.1103/PhysRevD.82.115025} {\bibfield  {journal} {\bibinfo  {journal}
  {Phys.Rev.},\ }\textbf {\bibinfo {volume} {D82}},\ \bibinfo {pages} {115025}
  (\bibinfo {year} {2010})},\ \Eprint {http://arxiv.org/abs/1010.1471}
  {arXiv:1010.1471 [hep-ph]} \BibitemShut {NoStop}%
\bibitem [{\citenamefont {Haba}\ and\ \citenamefont
  {Seto}(2011)}]{Haba:2011ra}%
  \BibitemOpen
  \bibfield  {author} {\bibinfo {author} {\bibfnamefont {N.}~\bibnamefont
  {Haba}}\ and\ \bibinfo {author} {\bibfnamefont {O.}~\bibnamefont {Seto}},\
  }\Doi {10.1143/PTP.125.1155} {\bibfield  {journal} {\bibinfo  {journal}
  {Prog.Theor.Phys.},\ }\textbf {\bibinfo {volume} {125}},\ \bibinfo {pages}
  {1155} (\bibinfo {year} {2011})}\BibitemShut {NoStop}%
\bibitem [{\citenamefont {Krauss}(1983)}]{Krauss:1983ik}%
  \BibitemOpen
  \bibfield  {author} {\bibinfo {author} {\bibfnamefont {L.~M.}\ \bibnamefont
  {Krauss}},\ }\Doi {10.1016/0550-3213(83)90574-6} {\bibfield  {journal}
  {\bibinfo  {journal} {Nucl.Phys.},\ }\textbf {\bibinfo {volume} {B227}},\
  \bibinfo {pages} {556} (\bibinfo {year} {1983})}\BibitemShut {NoStop}%
\bibitem [{\citenamefont {Cheung}\ \emph {et~al.}(2011)\citenamefont {Cheung},
  \citenamefont {Elor},\ and\ \citenamefont {Hall}}]{Cheung:2011mg}%
  \BibitemOpen
  \bibfield  {author} {\bibinfo {author} {\bibfnamefont {C.}~\bibnamefont
  {Cheung}}, \bibinfo {author} {\bibfnamefont {G.}~\bibnamefont {Elor}}, \ and\
  \bibinfo {author} {\bibfnamefont {L.~J.}\ \bibnamefont {Hall}},\ }\href@noop
  {} { (\bibinfo {year} {2011})},\ \Eprint {http://arxiv.org/abs/1104.0692}
  {arXiv:1104.0692 [hep-ph]} \BibitemShut {NoStop}%
\bibitem [{\citenamefont {Kribs}\ \emph {et~al.}(2008)\citenamefont {Kribs},
  \citenamefont {Poppitz},\ and\ \citenamefont {Weiner}}]{Kribs:2007ac}%
  \BibitemOpen
  \bibfield  {author} {\bibinfo {author} {\bibfnamefont {G.~D.}\ \bibnamefont
  {Kribs}}, \bibinfo {author} {\bibfnamefont {E.}~\bibnamefont {Poppitz}}, \
  and\ \bibinfo {author} {\bibfnamefont {N.}~\bibnamefont {Weiner}},\ }\Doi
  {10.1103/PhysRevD.78.055010} {\bibfield  {journal} {\bibinfo  {journal}
  {Phys.Rev.},\ }\textbf {\bibinfo {volume} {D78}},\ \bibinfo {pages} {055010}
  (\bibinfo {year} {2008})},\ \Eprint {http://arxiv.org/abs/0712.2039}
  {arXiv:0712.2039 [hep-ph]} \BibitemShut {NoStop}%
\bibitem [{\citenamefont {Lee}\ \emph {et~al.}(2011){\natexlab{a}}\citenamefont
  {Lee}, \citenamefont {Raby}, \citenamefont {Ratz}, \citenamefont {Ross},
  \citenamefont {Schieren} \emph {et~al.}}]{Lee:2010gv}%
  \BibitemOpen
  \bibfield  {author} {\bibinfo {author} {\bibfnamefont {H.~M.}\ \bibnamefont
  {Lee}}, \bibinfo {author} {\bibfnamefont {S.}~\bibnamefont {Raby}}, \bibinfo
  {author} {\bibfnamefont {M.}~\bibnamefont {Ratz}}, \bibinfo {author}
  {\bibfnamefont {G.~G.}\ \bibnamefont {Ross}}, \bibinfo {author}
  {\bibfnamefont {R.}~\bibnamefont {Schieren}},  \emph {et~al.},\ }\Doi
  {10.1016/j.physletb.2010.10.038} {\bibfield  {journal} {\bibinfo  {journal}
  {Phys.Lett.},\ }\textbf {\bibinfo {volume} {B694}},\ \bibinfo {pages} {491}
  (\bibinfo {year} {2011}{\natexlab{a}})},\ \Eprint
  {http://arxiv.org/abs/1009.0905} {arXiv:1009.0905 [hep-ph]} \BibitemShut
  {NoStop}%
\bibitem [{\citenamefont {Lee}\ \emph {et~al.}(2011){\natexlab{b}}\citenamefont
  {Lee}, \citenamefont {Raby}, \citenamefont {Ratz}, \citenamefont {Ross},
  \citenamefont {Schieren} \emph {et~al.}}]{Lee:2011dya}%
  \BibitemOpen
  \bibfield  {author} {\bibinfo {author} {\bibfnamefont {H.~M.}\ \bibnamefont
  {Lee}}, \bibinfo {author} {\bibfnamefont {S.}~\bibnamefont {Raby}}, \bibinfo
  {author} {\bibfnamefont {M.}~\bibnamefont {Ratz}}, \bibinfo {author}
  {\bibfnamefont {G.~G.}\ \bibnamefont {Ross}}, \bibinfo {author}
  {\bibfnamefont {R.}~\bibnamefont {Schieren}},  \emph {et~al.},\ }\Doi
  {10.1016/j.nuclphysb.2011.04.009} {\bibfield  {journal} {\bibinfo  {journal}
  {Nucl.Phys.},\ }\textbf {\bibinfo {volume} {B850}},\ \bibinfo {pages} {1}
  (\bibinfo {year} {2011}{\natexlab{b}})},\ \Eprint
  {http://arxiv.org/abs/1102.3595} {arXiv:1102.3595 [hep-ph]} \BibitemShut
  {NoStop}%
\bibitem [{\citenamefont {Siegel}\ and\ \citenamefont
  {Gates}(1979)}]{Siegel:1978mj}%
  \BibitemOpen
  \bibfield  {author} {\bibinfo {author} {\bibfnamefont {W.}~\bibnamefont
  {Siegel}}\ and\ \bibinfo {author} {\bibfnamefont {J.}~\bibnamefont {Gates},
  \bibfnamefont {S.James}},\ }\Doi {10.1016/0550-3213(79)90416-4} {\bibfield
  {journal} {\bibinfo  {journal} {Nucl.Phys.},\ }\textbf {\bibinfo {volume}
  {B147}},\ \bibinfo {pages} {77} (\bibinfo {year} {1979})}\BibitemShut
  {NoStop}%
\bibitem [{\citenamefont {Gates}\ \emph {et~al.}(1983)\citenamefont {Gates},
  \citenamefont {Grisaru}, \citenamefont {Rocek},\ and\ \citenamefont
  {Siegel}}]{Gates:1983nr}%
  \BibitemOpen
  \bibfield  {author} {\bibinfo {author} {\bibfnamefont {S.}~\bibnamefont
  {Gates}}, \bibinfo {author} {\bibfnamefont {M.~T.}\ \bibnamefont {Grisaru}},
  \bibinfo {author} {\bibfnamefont {M.}~\bibnamefont {Rocek}}, \ and\ \bibinfo
  {author} {\bibfnamefont {W.}~\bibnamefont {Siegel}},\ }\href@noop {}
  {\bibfield  {journal} {\bibinfo  {journal} {Front.Phys.},\ }\textbf {\bibinfo
  {volume} {58}},\ \bibinfo {pages} {1} (\bibinfo {year} {1983})},\ \Eprint
  {http://arxiv.org/abs/hep-th/0108200} {arXiv:hep-th/0108200 [hep-th]}
  \BibitemShut {NoStop}%
\bibitem [{\citenamefont {Kaplunovsky}\ and\ \citenamefont
  {Louis}(1994)}]{Kaplunovsky:1994fg}%
  \BibitemOpen
  \bibfield  {author} {\bibinfo {author} {\bibfnamefont {V.}~\bibnamefont
  {Kaplunovsky}}\ and\ \bibinfo {author} {\bibfnamefont {J.}~\bibnamefont
  {Louis}},\ }\Doi {10.1016/0550-3213(94)00150-2} {\bibfield  {journal}
  {\bibinfo  {journal} {Nucl.Phys.},\ }\textbf {\bibinfo {volume} {B422}},\
  \bibinfo {pages} {57} (\bibinfo {year} {1994})},\ \bibinfo {note} {dedicated
  to the memory of Brian Warr},\ \Eprint {http://arxiv.org/abs/hep-th/9402005}
  {arXiv:hep-th/9402005 [hep-th]} \BibitemShut {NoStop}%
\bibitem [{Note1()}]{Note1}%
  \BibitemOpen
  \bibinfo {note} {KR thanks Matthew McCullough for an insightful discussion on
  this point}\BibitemShut {NoStop}%
\bibitem [{\citenamefont {Hall}\ and\ \citenamefont
  {Randall}(1991)}]{Hall:1990hq}%
  \BibitemOpen
  \bibfield  {author} {\bibinfo {author} {\bibfnamefont {L.}~\bibnamefont
  {Hall}}\ and\ \bibinfo {author} {\bibfnamefont {L.}~\bibnamefont {Randall}},\
  }\Doi {10.1016/0550-3213(91)90444-3} {\bibfield  {journal} {\bibinfo
  {journal} {Nucl.Phys.},\ }\textbf {\bibinfo {volume} {B352}},\ \bibinfo
  {pages} {289} (\bibinfo {year} {1991})}\BibitemShut {NoStop}%
\bibitem [{\citenamefont {Randall}\ and\ \citenamefont
  {Rius}(1992)}]{Randall:1992cq}%
  \BibitemOpen
  \bibfield  {author} {\bibinfo {author} {\bibfnamefont {L.}~\bibnamefont
  {Randall}}\ and\ \bibinfo {author} {\bibfnamefont {N.}~\bibnamefont {Rius}},\
  }\Doi {10.1016/0370-2693(92)91779-9} {\bibfield  {journal} {\bibinfo
  {journal} {Phys.Lett.},\ }\textbf {\bibinfo {volume} {B286}},\ \bibinfo
  {pages} {299} (\bibinfo {year} {1992})}\BibitemShut {NoStop}%
\bibitem [{\citenamefont {Fox}\ \emph {et~al.}(2002)\citenamefont {Fox},
  \citenamefont {Nelson},\ and\ \citenamefont {Weiner}}]{Fox:2002bu}%
  \BibitemOpen
  \bibfield  {author} {\bibinfo {author} {\bibfnamefont {P.~J.}\ \bibnamefont
  {Fox}}, \bibinfo {author} {\bibfnamefont {A.~E.}\ \bibnamefont {Nelson}}, \
  and\ \bibinfo {author} {\bibfnamefont {N.}~\bibnamefont {Weiner}},\
  }\href@noop {} {\bibfield  {journal} {\bibinfo  {journal} {JHEP},\ }\textbf
  {\bibinfo {volume} {0208}},\ \bibinfo {pages} {035} (\bibinfo {year}
  {2002})},\ \Eprint {http://arxiv.org/abs/hep-ph/0206096}
  {arXiv:hep-ph/0206096 [hep-ph]} \BibitemShut {NoStop}%
\bibitem [{\citenamefont {Amigo}\ \emph {et~al.}(2009)\citenamefont {Amigo},
  \citenamefont {Blechman}, \citenamefont {Fox},\ and\ \citenamefont
  {Poppitz}}]{Amigo:2008rc}%
  \BibitemOpen
  \bibfield  {author} {\bibinfo {author} {\bibfnamefont {S.~D.~L.}\
  \bibnamefont {Amigo}}, \bibinfo {author} {\bibfnamefont {A.~E.}\ \bibnamefont
  {Blechman}}, \bibinfo {author} {\bibfnamefont {P.~J.}\ \bibnamefont {Fox}}, \
  and\ \bibinfo {author} {\bibfnamefont {E.}~\bibnamefont {Poppitz}},\ }\Doi
  {10.1088/1126-6708/2009/01/018} {\bibfield  {journal} {\bibinfo  {journal}
  {JHEP},\ }\textbf {\bibinfo {volume} {0901}},\ \bibinfo {pages} {018}
  (\bibinfo {year} {2009})},\ \Eprint {http://arxiv.org/abs/0809.1112}
  {arXiv:0809.1112 [hep-ph]} \BibitemShut {NoStop}%
\bibitem [{\citenamefont {Blechman}(2009)}]{Blechman:2009if}%
  \BibitemOpen
  \bibfield  {author} {\bibinfo {author} {\bibfnamefont {A.~E.}\ \bibnamefont
  {Blechman}},\ }\Doi {10.1142/S0217732309030795} {\bibfield  {journal}
  {\bibinfo  {journal} {Mod.Phys.Lett.},\ }\textbf {\bibinfo {volume} {A24}},\
  \bibinfo {pages} {633} (\bibinfo {year} {2009})},\ \Eprint
  {http://arxiv.org/abs/0903.2822} {arXiv:0903.2822 [hep-ph]} \BibitemShut
  {NoStop}%
\bibitem [{\citenamefont {Benakli}\ and\ \citenamefont
  {Goodsell}(2010)}]{Benakli:2010gi}%
  \BibitemOpen
  \bibfield  {author} {\bibinfo {author} {\bibfnamefont {K.}~\bibnamefont
  {Benakli}}\ and\ \bibinfo {author} {\bibfnamefont {M.}~\bibnamefont
  {Goodsell}},\ }\Doi {10.1016/j.nuclphysb.2010.06.018} {\bibfield  {journal}
  {\bibinfo  {journal} {Nucl.Phys.},\ }\textbf {\bibinfo {volume} {B840}},\
  \bibinfo {pages} {1} (\bibinfo {year} {2010})},\ \Eprint
  {http://arxiv.org/abs/1003.4957} {arXiv:1003.4957 [hep-ph]} \BibitemShut
  {NoStop}%
\bibitem [{\citenamefont {{Giudice}}\ and\ \citenamefont
  {{Masiero}}(1988)}]{1988PhLB..206..480G}%
  \BibitemOpen
  \bibfield  {author} {\bibinfo {author} {\bibfnamefont {G.~F.}\ \bibnamefont
  {{Giudice}}}\ and\ \bibinfo {author} {\bibfnamefont {A.}~\bibnamefont
  {{Masiero}}},\ }\Doi {10.1016/0370-2693(88)91613-9} {\bibfield  {journal}
  {\bibinfo  {journal} {Physics Letters B},\ }\textbf {\bibinfo {volume}
  {206}},\ \bibinfo {pages} {480} (\bibinfo {year} {1988})}\BibitemShut
  {NoStop}%
\bibitem [{\citenamefont {Davies}\ \emph {et~al.}(2011)\citenamefont {Davies},
  \citenamefont {March-Russell},\ and\ \citenamefont
  {McCullough}}]{Davies:2011mp}%
  \BibitemOpen
  \bibfield  {author} {\bibinfo {author} {\bibfnamefont {R.}~\bibnamefont
  {Davies}}, \bibinfo {author} {\bibfnamefont {J.}~\bibnamefont
  {March-Russell}}, \ and\ \bibinfo {author} {\bibfnamefont {M.}~\bibnamefont
  {McCullough}},\ }\Doi {10.1007/JHEP04(2011)108} {\bibfield  {journal}
  {\bibinfo  {journal} {JHEP},\ }\textbf {\bibinfo {volume} {1104}},\ \bibinfo
  {pages} {108} (\bibinfo {year} {2011})},\ \Eprint
  {http://arxiv.org/abs/1103.1647} {arXiv:1103.1647 [hep-ph]} \BibitemShut
  {NoStop}%
\bibitem [{Note2()}]{Note2}%
  \BibitemOpen
  \bibinfo {note} {One possibility would be if there existed $X'$ with $R$=-2
  and $F_{X'} \sim F_{X}$}\BibitemShut {NoStop}%
\bibitem [{\citenamefont {Davidson}\ \emph {et~al.}(2008)\citenamefont
  {Davidson}, \citenamefont {Nardi},\ and\ \citenamefont
  {Nir}}]{Davidson:2008bu}%
  \BibitemOpen
  \bibfield  {author} {\bibinfo {author} {\bibfnamefont {S.}~\bibnamefont
  {Davidson}}, \bibinfo {author} {\bibfnamefont {E.}~\bibnamefont {Nardi}}, \
  and\ \bibinfo {author} {\bibfnamefont {Y.}~\bibnamefont {Nir}},\ }\Doi
  {10.1016/j.physrep.2008.06.002} {\bibfield  {journal} {\bibinfo  {journal}
  {Phys.Rept.},\ }\textbf {\bibinfo {volume} {466}},\ \bibinfo {pages} {105}
  (\bibinfo {year} {2008})},\ \Eprint {http://arxiv.org/abs/0802.2962}
  {arXiv:0802.2962 [hep-ph]} \BibitemShut {NoStop}%
\bibitem [{\citenamefont {Sakharov}(1967)}]{Sakharov:1967dj}%
  \BibitemOpen
  \bibfield  {author} {\bibinfo {author} {\bibfnamefont {A.}~\bibnamefont
  {Sakharov}},\ }\href@noop {} {\bibfield  {journal} {\bibinfo  {journal}
  {Pisma Zh.Eksp.Teor.Fiz.},\ }\textbf {\bibinfo {volume} {5}},\ \bibinfo
  {pages} {32} (\bibinfo {year} {1967})}\BibitemShut {NoStop}%
\bibitem [{Note3()}]{Note3}%
  \BibitemOpen
  \bibinfo {note} {The next section explains why it is the physical Higgs
  }\BibitemShut {NoStop}%
\bibitem [{\citenamefont {Kolb}\ and\ \citenamefont
  {Turner}(1990)}]{Kolb:1990vq}%
  \BibitemOpen
  \bibfield  {author} {\bibinfo {author} {\bibfnamefont {E.~W.}\ \bibnamefont
  {Kolb}}\ and\ \bibinfo {author} {\bibfnamefont {M.~S.}\ \bibnamefont
  {Turner}},\ }\href@noop {} {\bibfield  {journal} {\bibinfo  {journal}
  {Front.Phys.},\ }\textbf {\bibinfo {volume} {69}},\ \bibinfo {pages} {1}
  (\bibinfo {year} {1990})}\BibitemShut {NoStop}%
\bibitem [{\citenamefont {Dine}\ \emph {et~al.}(1996)\citenamefont {Dine},
  \citenamefont {Nelson}, \citenamefont {Nir},\ and\ \citenamefont
  {Shirman}}]{Dine:1995ag}%
  \BibitemOpen
  \bibfield  {author} {\bibinfo {author} {\bibfnamefont {M.}~\bibnamefont
  {Dine}}, \bibinfo {author} {\bibfnamefont {A.~E.}\ \bibnamefont {Nelson}},
  \bibinfo {author} {\bibfnamefont {Y.}~\bibnamefont {Nir}}, \ and\ \bibinfo
  {author} {\bibfnamefont {Y.}~\bibnamefont {Shirman}},\ }\Doi
  {10.1103/PhysRevD.53.2658} {\bibfield  {journal} {\bibinfo  {journal}
  {Phys.Rev.},\ }\textbf {\bibinfo {volume} {D53}},\ \bibinfo {pages} {2658}
  (\bibinfo {year} {1996})},\ \Eprint {http://arxiv.org/abs/hep-ph/9507378}
  {arXiv:hep-ph/9507378 [hep-ph]} \BibitemShut {NoStop}%
\bibitem [{\citenamefont {Dine}\ \emph {et~al.}(1995)\citenamefont {Dine},
  \citenamefont {Nelson},\ and\ \citenamefont {Shirman}}]{Dine:1994vc}%
  \BibitemOpen
  \bibfield  {author} {\bibinfo {author} {\bibfnamefont {M.}~\bibnamefont
  {Dine}}, \bibinfo {author} {\bibfnamefont {A.~E.}\ \bibnamefont {Nelson}}, \
  and\ \bibinfo {author} {\bibfnamefont {Y.}~\bibnamefont {Shirman}},\ }\Doi
  {10.1103/PhysRevD.51.1362} {\bibfield  {journal} {\bibinfo  {journal}
  {Phys.Rev.},\ }\textbf {\bibinfo {volume} {D51}},\ \bibinfo {pages} {1362}
  (\bibinfo {year} {1995})},\ \Eprint {http://arxiv.org/abs/hep-ph/9408384}
  {arXiv:hep-ph/9408384 [hep-ph]} \BibitemShut {NoStop}%
\bibitem [{\citenamefont {Feng}\ \emph {et~al.}(1999)\citenamefont {Feng},
  \citenamefont {Moroi}, \citenamefont {Randall}, \citenamefont {Strassler},\
  and\ \citenamefont {Su}}]{Feng:1999fu}%
  \BibitemOpen
  \bibfield  {author} {\bibinfo {author} {\bibfnamefont {J.~L.}\ \bibnamefont
  {Feng}}, \bibinfo {author} {\bibfnamefont {T.}~\bibnamefont {Moroi}},
  \bibinfo {author} {\bibfnamefont {L.}~\bibnamefont {Randall}}, \bibinfo
  {author} {\bibfnamefont {M.}~\bibnamefont {Strassler}}, \ and\ \bibinfo
  {author} {\bibfnamefont {S.-f.}\ \bibnamefont {Su}},\ }\Doi
  {10.1103/PhysRevLett.83.1731} {\bibfield  {journal} {\bibinfo  {journal}
  {Phys.Rev.Lett.},\ }\textbf {\bibinfo {volume} {83}},\ \bibinfo {pages}
  {1731} (\bibinfo {year} {1999})},\ \Eprint
  {http://arxiv.org/abs/hep-ph/9904250} {arXiv:hep-ph/9904250 [hep-ph]}
  \BibitemShut {NoStop}%
\bibitem [{\citenamefont {Raby}(1997)}]{Raby:1997pb}%
  \BibitemOpen
  \bibfield  {author} {\bibinfo {author} {\bibfnamefont {S.}~\bibnamefont
  {Raby}},\ }\Doi {10.1103/PhysRevD.56.2852} {\bibfield  {journal} {\bibinfo
  {journal} {Phys.Rev.},\ }\textbf {\bibinfo {volume} {D56}},\ \bibinfo {pages}
  {2852} (\bibinfo {year} {1997})},\ \Eprint
  {http://arxiv.org/abs/hep-ph/9702299} {arXiv:hep-ph/9702299 [hep-ph]}
  \BibitemShut {NoStop}%
\bibitem [{\citenamefont {Baer}\ \emph {et~al.}(1999)\citenamefont {Baer},
  \citenamefont {Cheung},\ and\ \citenamefont {Gunion}}]{Baer:1998pg}%
  \BibitemOpen
  \bibfield  {author} {\bibinfo {author} {\bibfnamefont {H.}~\bibnamefont
  {Baer}}, \bibinfo {author} {\bibfnamefont {K.-m.}\ \bibnamefont {Cheung}}, \
  and\ \bibinfo {author} {\bibfnamefont {J.~F.}\ \bibnamefont {Gunion}},\ }\Doi
  {10.1103/PhysRevD.59.075002} {\bibfield  {journal} {\bibinfo  {journal}
  {Phys. Rev.},\ }\textbf {\bibinfo {volume} {D59}},\ \bibinfo {pages} {075002}
  (\bibinfo {year} {1999})},\ \Eprint {http://arxiv.org/abs/hep-ph/9806361}
  {arXiv:hep-ph/9806361} \BibitemShut {NoStop}%
\bibitem [{\citenamefont {Giudice}\ and\ \citenamefont
  {Romanino}(2004)}]{Giudice:2004tc}%
  \BibitemOpen
  \bibfield  {author} {\bibinfo {author} {\bibfnamefont {G.}~\bibnamefont
  {Giudice}}\ and\ \bibinfo {author} {\bibfnamefont {A.}~\bibnamefont
  {Romanino}},\ }\Doi {10.1016/j.nuclphysb.2004.11.048,
  10.1016/j.nuclphysb.2004.11.048} {\bibfield  {journal} {\bibinfo  {journal}
  {Nucl.Phys.},\ }\textbf {\bibinfo {volume} {B699}},\ \bibinfo {pages} {65}
  (\bibinfo {year} {2004})},\ \Eprint {http://arxiv.org/abs/hep-ph/0406088}
  {arXiv:hep-ph/0406088 [hep-ph]} \BibitemShut {NoStop}%
\bibitem [{\citenamefont {Arkani-Hamed}\ and\ \citenamefont
  {Dimopoulos}(2005)}]{ArkaniHamed:2004fb}%
  \BibitemOpen
  \bibfield  {author} {\bibinfo {author} {\bibfnamefont {N.}~\bibnamefont
  {Arkani-Hamed}}\ and\ \bibinfo {author} {\bibfnamefont {S.}~\bibnamefont
  {Dimopoulos}},\ }\Doi {10.1088/1126-6708/2005/06/073} {\bibfield  {journal}
  {\bibinfo  {journal} {JHEP},\ }\textbf {\bibinfo {volume} {0506}},\ \bibinfo
  {pages} {073} (\bibinfo {year} {2005})},\ \Eprint
  {http://arxiv.org/abs/hep-th/0405159} {arXiv:hep-th/0405159 [hep-th]}
  \BibitemShut {NoStop}%
\bibitem [{\citenamefont {Bagger}\ \emph {et~al.}(1995)\citenamefont {Bagger},
  \citenamefont {Poppitz},\ and\ \citenamefont {Randall}}]{Bagger:1995ay}%
  \BibitemOpen
  \bibfield  {author} {\bibinfo {author} {\bibfnamefont {J.}~\bibnamefont
  {Bagger}}, \bibinfo {author} {\bibfnamefont {E.}~\bibnamefont {Poppitz}}, \
  and\ \bibinfo {author} {\bibfnamefont {L.}~\bibnamefont {Randall}},\ }\Doi
  {10.1016/0550-3213(95)00463-3} {\bibfield  {journal} {\bibinfo  {journal}
  {Nucl.Phys.},\ }\textbf {\bibinfo {volume} {B455}},\ \bibinfo {pages} {59}
  (\bibinfo {year} {1995})},\ \Eprint {http://arxiv.org/abs/hep-ph/9505244}
  {arXiv:hep-ph/9505244 [hep-ph]} \BibitemShut {NoStop}%
\bibitem [{\citenamefont {Abel}\ and\ \citenamefont
  {Goodsell}(2011)}]{Abel:2011dc}%
  \BibitemOpen
  \bibfield  {author} {\bibinfo {author} {\bibfnamefont {S.}~\bibnamefont
  {Abel}}\ and\ \bibinfo {author} {\bibfnamefont {M.}~\bibnamefont
  {Goodsell}},\ }\Doi {10.1007/JHEP06(2011)064} {\bibfield  {journal} {\bibinfo
   {journal} {JHEP},\ }\textbf {\bibinfo {volume} {1106}},\ \bibinfo {pages}
  {064} (\bibinfo {year} {2011})},\ \Eprint {http://arxiv.org/abs/1102.0014}
  {arXiv:1102.0014 [hep-th]} \BibitemShut {NoStop}%
\end{thebibliography}
%
\end{document}